\documentclass[aps,numerical,graphicx,reprint,pre]{revtex4-1}
\usepackage{graphicx}
\draft % marks overfull lines with a black rule on the right

\begin{document}

%\preprint{}
% A useful Journal macro 
\def\Journal#1#2#3#4{{#1 }{\bf #2, }{ #3 }{ (#4)}} 
 
% Some useful journal names 
\def\BiJ{ Biophys. J.}                 
\def\Bios{ Biosensors and Bioelectronics} 
\def\LNC{ Lett. Nuovo Cimento} 
\def\JCP{ J. Chem. Phys.} 
\def\JAP{ J. Appl. Phys.} 
\def\JMB{ J. Mol. Biol.} 
\def\JPC{ J. Phys: Condens. Matter}
\def\CMP{ Comm. Math. Phys.} 
\def\LMP{ Lett. Math. Phys.} 
\def\NLE{{ Nature Lett.}} 
\def\NPB{{ Nucl. Phys.} B} 
\def\PLA{{ Phys. Lett.}  A} 
\def\PLB{{ Phys. Lett.}  B} 
\def\PNAS{Proc. Natl. Am. Soc.}
\def\PRL{ Phys. Rev. Lett.} 
\def\PRA{{ Phys. Rev.} A} 
\def\PRE{{ Phys. Rev.} E} 
\def\PRB{{ Phys. Rev.} B} 
\def\PD{{ Physica} D} 
\def\ZPC{{ Z. Phys.} C} 
\def\RMP{ Rev. Mod. Phys.} 
\def\EPJD{{ Eur. Phys. J.} D} 
\def\SAB{ Sens. Act. B} 
%%%%%%%%%%%%%%%%%%%%%%%%%%%%%%%%%%%%%
\title{On the mechanisms responsible of photocurrent in bacteriorhodopsin 
%$or$ \\ global and local effects of photon interaction with bR
}

\author{Eleonora Alfinito}
\email{eleonora.alfinito@unisalento.it}
\homepage{http://cmtg1.unile.it/eleonora1.html}
\affiliation{Dipartimento di Ingegneria dell'Innovazione,
Universit\`a del Salento, via Monteroni, I-73100 Lecce, Italy}
\affiliation{CNISM,  Via della Vasca Navale, 84 - 00146 Roma, Italy }
\author{Lino Reggiani}
%\thanks{ovvero?}
\email{lino.reggiani@unisalento.it}
%\thanks{}
\affiliation{Dipartimento di Matematica e Fisica, "Ennio de Giorgi",
Universit\`a del Salento, via Monteroni, I-73100 Lecce, Italy}
\affiliation{CNISM,  Via della Vasca Navale, 84 - 00146 Roma, Italy}

\date{\today}
\begin{abstract}
Recently, a growing interest has been addressed to  the electrical properties of  bacteriorhodopsin (bR), a protein belonging to the transmembrane protein family. 
Several experiments pointed out the role of green light in enhancing the current flow in nanolayers of bR, thus confirming potential applications of this protein in the field of optoelectronics .
By contrast, the mechanisms underlying the charge transfer and the associated photocurrent are still far from being understood at a microscopic level.
To take into account  the structure-dependent nature of the current, in a previous set of papers we suggested a mechanism of sequential tunneling among neighbouring amino acids.
As  a matter of fact, it is well accepted that, when  irradiated with green light, bR undergoes a conformational change at a molecular level.
Thus, the role played by the protein tertiary-structure in modeling the charge transfer cannot be neglected. 
The aim of this paper is to go beyond previous models, in the framework of a new branch of electronics, we called proteotronics, which exploits the ability to use proteins as reliable, well understood materials,  for the development of novel bioelectronic devices. 
In particular, the present approach assumes that the conformational change is not the unique transformation that the protein undergoes when irradiated by light.
Instead, the light can also promote a free-energy increase of the protein state  that, in turn, should  modify its internal degree of connectivity, here described by the change  in the value of an interaction radius associated with the physical interactions among amino acids.
The implemented model enables us to achieve a better agreement between theory and experiments in the region of a low applied bias by preserving the level of agreement at high values of applied bias.
Furthermore, results provide new insights on the mechanisms responsible for bR photoresponse.
\end{abstract}


\maketitle %\maketitle must follow title, authors, abstract and \pacs

\section{Introduction}
\label{Introduction}
Bacteriorhodopsin (bR) is the best known protein in the family of opsins, proteins conjugated with a molecule of retinal and able  to convert  visible light into electrostatic  energy \cite{Lozier75}.
This protein is found in a primeval organism, the \textit{Halobacterium salinarum}, specifically in a part of its cell membrane called purple membrane (PM), since its color. 
This membrane, $5 \  nm$ thick, is a natural thin film, essentially constituted by few lipids and these proteins organized in an hexagonal lattice \cite{Corcelli00}.
\par
A large number of studies has been carried out on bR in the field of biophysics and physicochemistry \cite{varo99,etzkorn13,yoshino13}, and many aspects have been unveiled.
As relevant examples we cite: 
(i) the photoinduced isomerization of the retinal embedded in bR \cite{Gonzales00}, 
(ii) the conformational change of bR associated with the retinal isomerization \cite{subramaniam00},
(iii)  the importance of environmental conditions in the photocycle development \cite{Luecke00,Kouyama89}.
\par
Patches of PM have been used for several purposes: to produce metal-protein-metal junctions \cite{Jin06,Ron10,Sepunaru12},  to perform c-AFM investigations\cite{Casuso07,Mukho14},  to develop solar cells of new generation \cite{Renu14}, etc.
As a matter of fact, films of bR resist to thermal, electrical and also mechanical stress \cite{Jin06,Casuso07,Ron10,Sepunaru12,Mukho14} and show a substantial photocurrent when irradiated by a visible (green) light \cite{Jin06,Sepunaru12,patil12}.
Therefore, bR can be used as an optoelectrical switch,  to convert radiant energy into electrical energy \cite{Renu14}, in pollutants remediation systems  \cite{Dai14}, to produce optical memories \cite{Hampp00}, to control neuronal and tissue activity \cite{Lima05,Deisseroth10},  etc.
\par
The commonly accepted view concerning the protein activation is the following:
A photon is absorbed by the retinal molecule contained in each protein, then causing the bending of this molecule. 
As a consequence, the protein undergoes a change of its tertiary structure, following a cycle of transformations that arrives to release a proton outside the cell membrane. Finally,  reprotonization of the retinal molecule by Asp96 restores  the native configuration.
Some crystallographic investigations have been performed on this protein to determine its configuration in the different steps of the cycle \cite{Luecke00}.    
This is a particularly hard task, since the X-ray radiation could modify the protein structure, and only recently the puzzle of many contradictory results starts to be recomposed \cite{Wickstrand14}.
At present, a rather complete description of the protein is given only for the native and the active L-state \cite{Berman00,Lanyi06}.
\par
Measurements of the protein current-voltage (I-V) characteristics were reported in several papers \cite{Jin06,Casuso07,Ron10,Sepunaru12,Mukho14}.
To this purpose, samples made of patches of PM were anchored on a conductive substrate and  connected to an external circuit. The connection was  made with : i) an extended transparent conductive contact \cite{Jin06,Ron10,Sepunaru12}, ii)  a tip of a c-AFM \cite{Casuso07,Mukho14}. 
In both  cases, the measured current was found to be quasi-Ohmic at the lowest bias, and strongly super-linear at increasing bias. 
Furthermore,  when the sample was irradiated with green light \cite{Jin06,Sepunaru12}, a significant photocurrent was observed.
There was a clear proof that the charge transfer is mainly  due to the protein \cite{Jin06, Sepunaru12}.
There was also a high resistance channel due to the lipid membrane, which is detected in experiments  involving a membrane deformation \cite{Alfinito11c,Casuso07}.
In the absence of membrane deformation, this channel can be neglected.
The charge transfer through the protein was attributed to a tunneling mechanism \cite{Jin06, Casuso07,Alfinito09,Alfinito11c}.
In particular, the presence of a current (in dark) well above possible leakage components, and of  a photocurrent (in light) supports the hypothesis of a mechanism of charge transfer intrinsically dependent on the protein tertiary structure \cite{Alfinito11c,Alfinito09,Alfinito09e,Alfinito09d,Nadia,Alfinito13b,Alfinito13c, Alfinito14a,Alfinito14b}. 
\par
Since a long  time,  the interaction of electromagnetic fields with biological matter is the object of many investigations, mainly for the damages produced by ionizing radiations.
As far as known, sunlight that reaches the Earth is largely composed of non-ionizing radiations whose main effect on biological matter is heating.
In particular, for proteins, this should lead to a global energy enhancement, regardless of the protein specific conformational state, as confirmed by recent experiments  showing the critical role of temperature in current measurements \cite{Sepunaru12}.
Therefore, we conjecture that in a sample of proteins, like a patch of purple membrane, light gives rise to different effects.
From one side, the retinal modification with the consequent conformational change from the native to the active state, from another side, a net transfer of energy to the whole protein with a consequent increase of its free-energy. 
%
%$^2$
%
As a general issue, both these effects should contribute to the protein activation.
%Indeed, it has been recently shown that, like in semiconductors (better if extrinsic), a  small amount of radiant energy could be sufficient to produce a surplus of conduction \cite{Sepunaru12}.
\par
The present paper addresses this issue by  accounting simultaneously for these two effects in a computational/theoretical model called INPA (impedance network protein analogue). 
This approach describes the electrical characteristics of a protein by using a network of impedances.
% and has been previously developed and used to model the results of some available experiments \cite{Jin06,Casuso07}.
In previous investigations, the local interaction of a photon with the retinal has been investigated by considering the corresponding change of the network structure
%, from 
%ONE ASSOCIATED WITH the protein in its native state, to anOther ASSOCIATED WITH the protein in an active state. 
The novelty of the present paper consists in the further introduction of a global energy increase of the the whole protein due to the incident light.
This is described by a change of the network connectivity both of the native and the active state.

\par
The methodological approach we follow points to the integration of  
different disciplines (molecular biology, physics, electronics) to develop a new generation of electronic devices within a nano-bio-technology.
This interdisciplinary approach is leading to an entirely new discipline which we christen \textit{proteotronics} \cite{rome14}.
\par
The paper is organized as follows.
Section II summarizes the main steps of the INPA model and describes the improvements introduced on the basis of the dynamical evolution of the protein energy landscape. 
Section III reports and discusses the main results and suggests the opening of new perspectives.  
Major conclusions are summarized in Section IV. 
\section {Theoretical model} 
%
%sec_2
The INPA model is based on a percolative approach that describes the protein like a network of links and nodes. 
A node represents a single amino acid and its spatial position is the same of the corresponding  $C_{\alpha}$ atom. 
A link joins a couple of nodes, and represents the interaction between amino acids\cite{Alfinito05,Akimov06,Alfinito08}.  
The protein structure in its native or active state is taken by public databases or homology modeling \cite{Berman00,Alfinito09a}, 
thus the  node configuration reproduces the protein backbone.
Then, couples of nodes are connected with the rule that they must be be closer than an assigned interaction radius, $R_{c}$.
In this way, the number of links, $N$, depends on the value of $R_c$ and is in the range $0  \le N_{aa} \le N_{aa}(N_{aa}-1)/2$, with $N_{aa}$ the number of amino acids pertaining to the given protein. 
In the present case, the macroscopic quantity of interest is the static I-V characteristic.
Therefore, the network is drawn like an electrical circuit where an elementary resistance, $R_{i,j}$, is associated with each link  between nodes $i$ and $j$.  
Explicitly:
\begin{equation}
R_{i,j}=\frac{l_{i,j}}{{\cal A}_{i,j}} \rho
\label{eq:1}                                                            
\end{equation}
where ${\cal A}_{i,j}\,=\,\pi (R_c^{2}-l_{i,j}/4)$, is the cross-sectional area between two spheres of radius $R_c$ centered on the $i$-th  and $j$-th node, respectively; $l_{i,j}$ is the distance between the sphere centers, $\rho$ is the resistivity. 
\par
By positioning the input and output electrical contacts, respectively, on the first and last node (more structured contacts can be envisioned)  for a given applied bias (current or voltage operation modes according to convenience) the network is solved within a linear Kirchhoff scheme and its global resistance, $R$,  is calculated  \cite{rome14, Alfinito11c,Alfinito09,Alfinito09e,Alfinito09d,Nadia,Alfinito13b,Alfinito13c,Alfinito14a,Alfinito14b}.
Accordingly, this network produces a parameter-dependent static I-V characteristic for the single protein, based on the standard relation:    
\begin{equation}
V = R I. 
\label{eq:2}
\end{equation}
To account for the super-linear behaviour of current at increasing voltages, a tunneling mechanism of charge transfer is included.
In doing so,  a stochastic approach within a Monte Carlo scheme \cite{Alfinito09d, Alfinito09, Alfinito11c, Alfinito13b, Alfinito13c, Alfinito14a, Alfinito14b} is used. 
In particular, following the Simmons model \cite{Simmons63}, a mechanism containing two possible tunneling processes, a direct tunneling (DT) at low bias, and a Fowler-Nordheim tunneling (FN) at high bias,  is introduced. 
Therefore, the resistivity value of each link is chosen between a low value $\rho_{\rm min}$, taken to fit the current at the highest voltages, and a high value $\rho(V)$, which depends on the voltage drop between network nodes as:
\begin{eqnarray}
\rho(V)&=& \rho_{\rm MAX}  \qquad \qquad (eV < \Phi),  \label{eq:3a}\\
\rho(V)&=&\rho_{\rm MAX}(\frac{\Phi}{eV})+\rho_{\rm min}(1- \frac{\Phi}{eV}) \quad   (eV \ge  \Phi) 
\label{eq:3b}
\end{eqnarray}
where $\rho_{\rm MAX}$ is the maximal resistivity value taken to fit the I-V characteristic at the lowest voltages (Ohmic response)  and $\Phi$  is the height of the tunneling barrier between nodes.
The transmission probability of each tunneling process is given by \cite{Alfinito09d, Alfinito09}:
\begin{eqnarray}
 P^{\rm DT}_{i,j} &=& \exp \left[- \alpha \sqrt{(\Phi-\frac{1}{2}
eV_{i,j})} \right] \quad  
 (eV_{i,j}  < \Phi) , \label{eq:4a} \\
{P}^{\rm FN}_{ij}&=&\exp \left[-\alpha\ \frac{\Phi}{eV_{i,j}}\sqrt{\frac{\Phi}{2}} \right] 
 \qquad   (eV_{i,j} \ge \Phi)  
\label{eq:4b}
\end{eqnarray}
where $V_{i,j}$ is the potential drop between the couple of $i,j$ amino acids, $\alpha =\frac{2l_{i,j}\sqrt{2m}}{\hbar}$, 
%$l_{i,j}$ is the distance between the $i$-th and $j$-th node 
and $m$  is the electron effective mass, here taken the same of the bare value.
The DT  superscript refers to the low-bias, quasi-Ohmic response and the FN subscript refers to the high-bias, super-Ohmic response.
%
%%%%%%%%%%%%%%%%%
\section{Results and Discussion}
%
%sec_3
By construction, both the current response at very low and very high bias exhibit an Ohmic behaviour with values of the corresponding  resistance differing for several orders of magnitude.
This model was successfully used to reproduce the experiments of Ref.~\cite{Casuso07}. 
The inputs parameters were $R_c$=6 \ \AA,\ $\Phi$=219 meV,   
$\rho_{MAX}=4 \times 10^{13} \ \Omega$ \AA, \  for the low field resistivity and  $\rho_{min}= 4 \times 10^{4} \ \Omega$\AA \ for the high field resistivity.
The protein tertiary structure was taken from the protein database \cite{Berman00}, specifically the 2NTU entry, an X-ray crystallographic measurement for bR  native-state.
\par
\begin{figure}
        \centering
                \includegraphics*[width=0.45\textwidth]{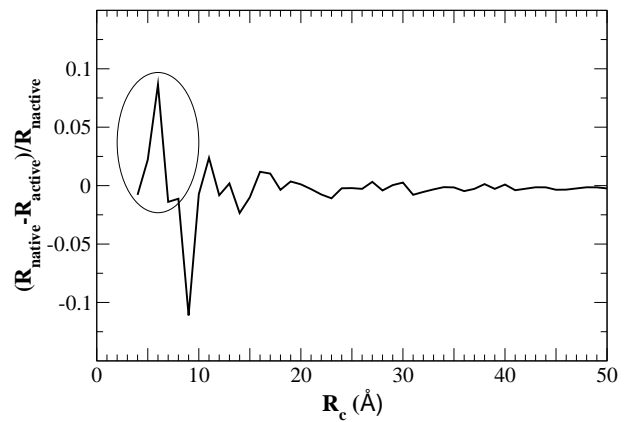}
        \caption{Relative resistance variations vs the interaction radius, $R_c$, for bR in the native and active state. The ellipse indicates the region of $R_c$ values whose trend is in agreement with experiments.}
        \label{fig:rel_res}
\end{figure}

The agreement between calculations and experiments was found to be satisfactory, also reproducing the current modifications due to the membrane indentation by the c-AFM tip \cite{Alfinito11c}.
On this basis, we found reasonable to take the same input to fit the response of the  protein in light.
At present, the only crystallographic entry describing the complete protein in an active state is 2NTW, which gives account of the L-state (henceforth called the active state) of bR. 
This state is sensitive to the $550 \ nm$ light and precedes the M state ($410 \ nm$), which corresponds to a proton releasing.
In the I-V measurements, the proton releasing was not monitored and the current measured was only attributed to electron transfer \cite{Jin06}.
When the activated configuration was used as input to fit the current response in the presence of light, the agreement with experiments was less satisfactory than that in dark.
A possible way to overcome this drawback is to assume that the presence of light modifies not only the protein structure but also its connectivity properties.
In the INPA model this modification is accounted for by changing the value of
the interaction radius.
To this purpose, Fig.~\ref{fig:rel_res} reports the role of the interaction radius in the calculation of the resistances of the native and active states.
Numerical data are obtained at very low voltages, where the Ohmic regime strictly holds, and reflect the protein topology.
%
%figure 1
The main results of Fig.~\ref{fig:rel_res} are: (i) the general low resolution between these states and, (ii) the presence of two regions in which the resolution is best appreciable, around  $R_c= 6 $ \ \AA \ and $R_c= 9 $ \ \AA. 
The experiments are in agreement with $R_c= 6 $ \ \AA.
\par
In the following the protein current responses are numerically analyzed for several values of $R_c$ around 6 \ \AA.
%figure 2
\begin{figure}
        \centering
                \includegraphics*[width=0.45\textwidth]{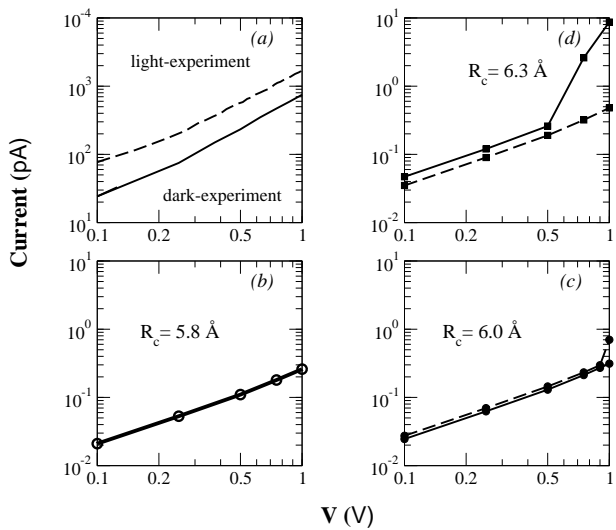}
        \caption{I-V characteristics of bR. 
Panel (a) reports the  experiments carried out on nanolayers samples \cite{Jin06}             Panels (b,c,d) refer to data calculated within the INPA model  (single protein) for the different values of $R_c$ reported in the figures. Symbols refer to numerical  calculations, lines are guides to the eyes.                            
Dashed lines and superimposed symbols refer to the active state; continuous lines and superimposed symbols to the native state.     
For $R_c$=5.8 \ \AA \ the I-V characteristics are found to coincide  for native and active states.}
        \label{fig:puri}
\end{figure}
%
%figure3
\begin{figure}
        \centering
                \includegraphics*[width=0.45\textwidth]{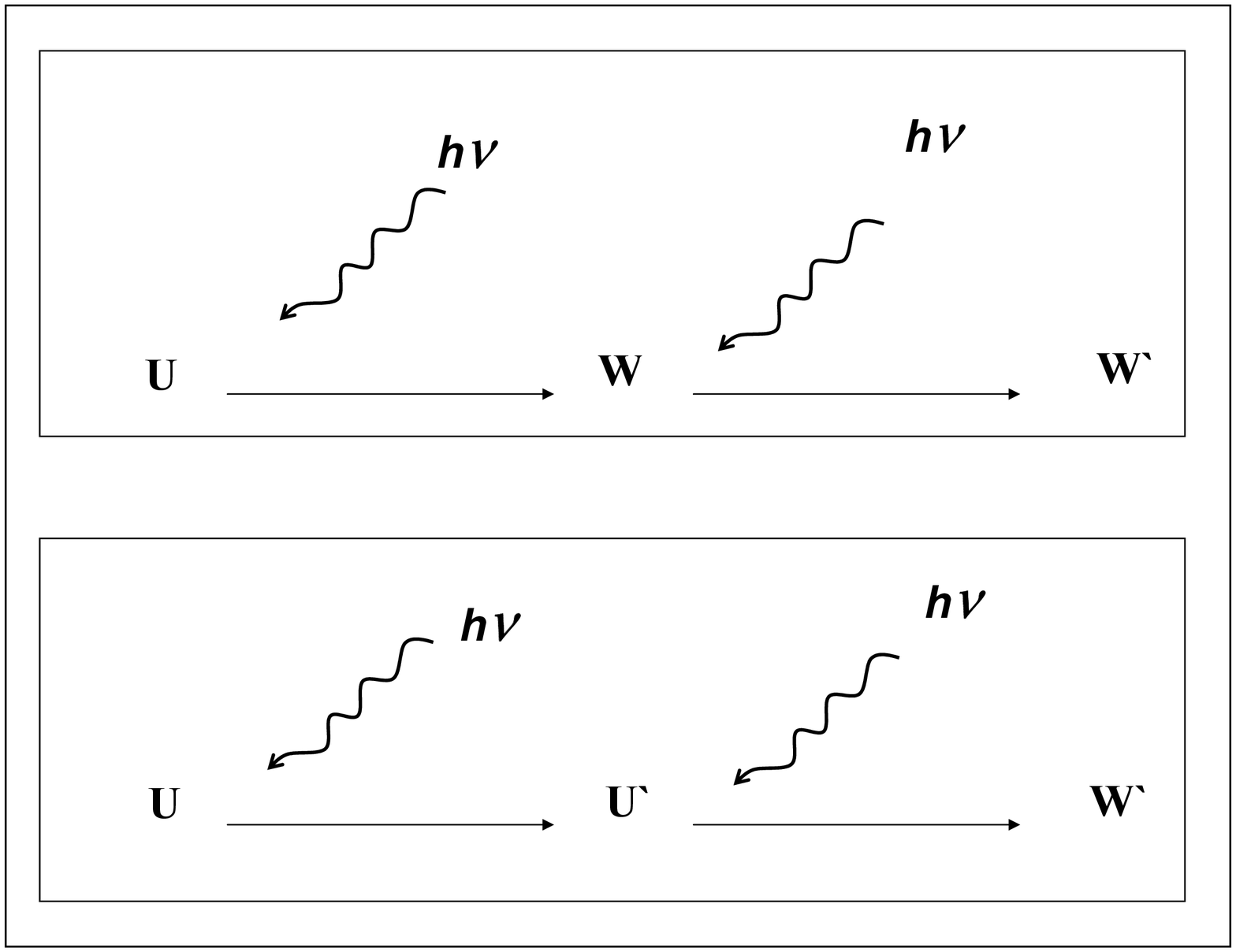}
        \caption{Schematic representation  of the  energy evolution by absorption of  photons  in a bR molecule. 
Upper panel depicts the conformational change from the native state U to an active state W induced by the absorption of a photon by the retinal and a successive global energy increase from W to W$^{\prime}$ induced by the absorption (from the whole protein) of other photons .
Lower panel depicts an alternative possibility when the absorption of  photons  induces a global energy increase from U to U$^{\prime}$ of the native state and a successive absorption process induce a conformational change from U$^{\prime}$ to the energy level W$^{\prime}$ of the active state.}
        \label{fig:schemaattivazione}
\end{figure}
%
%figure 4
\begin{figure}
        \centering
                \includegraphics*[width=0.45\textwidth]{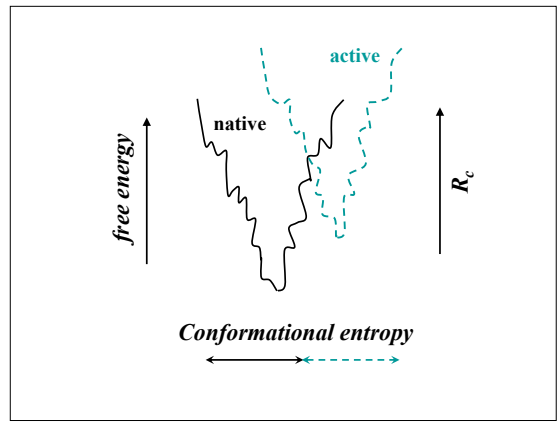}
        \caption{Schematic representation of the free-energy funnel landscape for native and active states. The minima of the free-energy corresponds to a minimum of the conformational entropy that measures the number of available microscopic states.
The increase of connectivity at increasing free-energy is depicted by the increase of the interaction radius $R_c$.}
        \label{fig:funnels}
\end{figure}
%
%figure5
\begin{figure}
        \centering
                \includegraphics*[width=0.45\textwidth]{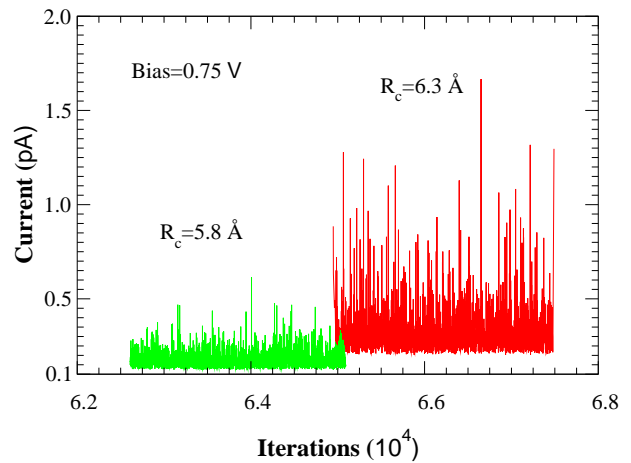}
        \caption{Current fluctuations of bR in its active state at different values of $R_c$}
        \label{fig:fluctuations}
\end{figure}
In particular, simulations for the single protein are performed for the three values, $R_{c}$= 5.8, 6.0 and 6.3 \ \AA\, and for both native and active state. 
Results are reported in Fig.~\ref{fig:puri}  with panel (a) reporting  the experimental data carried out in a bR macroscopic sample \cite{Jin06} in dark and light. 
For a given protein state,  Fig.~\ref{fig:puri} shows a current enhancement by increasing $R_c$ from 5.8 to 6.3 \ \AA.
Furthermore, at increasing $R_c$, the differences between the current response in dark and light are  more and more marked.
\par
The above results  suggest that the activation mechanism of a \textit{macroscopic sample} of bR can be  described within the \textit{single} protein model by using: i) a conformational change (from 2NTU to 2NTW structure), ii)  a connectivity change (i.e. a variation of the network interaction radius).
\par
More specifically, we can envision a twofold mechanism of photon absorption:
by the retinal, and by the whole protein.
The former is responsible of the conformational change, the latter of a global
energy increase of the protein. 
Notice that, according with experiments \cite{Sepunaru12}, the global energy increase is coherently used by the protein sample in enhancing its photocurrent response.
In other words, the  electromagnetic radiation impinging on the protein may anyway produce the global effect of an  energy gain, while  the local interaction of a photon with the retinal triggers the conformational change when the protein is in its native state. 
These mechanisms associated with photon absorption are schematically depicted in Fig.~\ref{fig:schemaattivazione}. 
From one hand, when the native state, say $U$, becomes an active state, say $W$, a further irradiation should  enhance the global energy of the active state.
In this way, the active state is promoted to an upper energy value $W^{\prime}$.
From another hand, when the $U$ state does not undergo a conformational change, anyway its energy level can be promoted to an upper value $U^{\prime}$; a further dose of light may drive this state to an active state $W^{\prime}$. 
\par
Among the different ways used to  describe the protein energy landscape at different stages of the folding, one of the most accepted is the rugged funnel-diagram \cite{Onuchic97}.
In this diagram, the protein folds from the molten state to the native (stable) state following many possible folding routes toward the minimum of a funnel-like energy  surface. 
When the protein runs down in the energy funnel, it loses the spurious bonds and enforces those stabilizing the minimal-energy configuration \cite{Onuchic97}. 
In doing so, it also reaches the minimum of the configurational entropy.
Furthermore, the phase transition from a stable state at low energy to a stable state at higher energy is depicted in terms of a tunneling between the minima of a multiwell energy landscape \cite{Kobilka07}.
%Figure \ref{fig:funnels} illustrates 
%such energy enhancement of the protein energy state due to light, which also produces an entropy enhancement.
As energy increases, the spurious connections do again appear and the protein can explore more microstates. 
In a very schematic way, this mechanism is pictured in Fig.~\ref{fig:funnels} where a couple of funnels representing the native and an active state are superimposed. 
The conformational change corresponds to the transition from a funnel to another one.
The minimal energy between the two funnel stable minima is of the order of the eV.
Otherwise, by rising the energy of the protein in the native state, it is possible to reach the overlapping region of the two funnels. 
Here the transition from the native to the active state can occur without energy supply.
\par 
Within the INPA model, the mechanism of energy increase is described by an increase of the $R_c$ value.
As a consequence, the network becomes more connected which implies an increase of the pathways for tunneling and of the number of possible current channels.
This, in turn,  leads to an increase of the instantaneous current fluctuations,  as reported in Fig.~\ref{fig:fluctuations}.
Here, the current fluctuations observed from simulations are reported for the active state, with an applied bias of 0.75 V and for two $R_{c}$ values of 5.8 and 6.3 \ \AA.
\par
Following this scheme, the current response of  samples made by monolayers of bR has been fitted by using a binary mixture of native and active states; the percentages of each state being a function of the $R_c$ value \cite{Nadia}.
Specifically, a good fit of the experimental data  \cite{Jin06} is obtained by using: 
(i) for the sample in light, $R_c$ = 6.3 \ \AA \ and a binary mixture of 96\% of 2NTW and a 4\% of 2NTU;
(ii) For the sample in dark $R_c$ = 5.8  \ \AA\ and 100\% of 2NTU.
\par
The c-AFM experiment, performed in the absence of direct light \cite{Casuso07},  was previously fitted within a very good accuracy on the full bias range by using $R_c$ = 6.0 \ \AA\ and 100\% of the 2NTW native state \cite{Alfinito11c}.  
Since in these experiments one cannot exclude the presence of a certain amount of proteins IN the active state, in agreement with a value of $R_c$ larger than the threshold value $R_c$ = 5.8 \ \AA\,
here  the fit with experiments is tested  by using  binary mixtures with an increasing percentage of active states. 
The fit is found to be sufficiently accurate with a percentage of active state not larger than 40\%.
%
%figureactivation
%figure 6
%
\begin{figure}
        \centering
                \includegraphics*[width=0.45\textwidth]{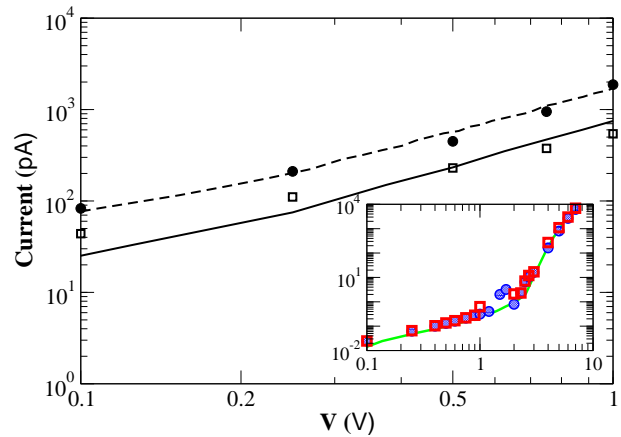}
        \caption{I-V characteristics calculated by using different mixtures of bR native and active states. Full circles refer to the mixture  4\% of native states and 96\% of active states with $R_c$=6.3 \ \AA. Open squares refer to a pure native state with $R_c$=5.8 \ \AA. Continuous (dashed) line refers to experimental data \cite{Jin06} in dark (light)  in the bias range $0.1 \div 1$ V.   In the inset the continuous line refers to experimental data in dark \cite{Casuso07}, in the bias range $0.1 \div 10$ V. Circles refer to data calculated with the pure native state with $R_c$=6.0 \ \AA, squares refer to data calculated with the mixture of 60\% of native states and 40\% of active states, with  $R_c$=6.0  \ \AA.  }
\label{fig:miscele}
\end{figure}
Figure \ref{fig:miscele} reports: (i) the experimental data \cite{Jin06, Casuso07} and,  
(ii) the single protein data rescaled by using the formula 
\begin{equation}
I_{S}=A\left(I_{n} \cdot n_{n}+I_{a} \cdot n_{a}\right)
\label{eq:5} 
\end{equation}
where $I_{S}$ indicates the sample current,  $A$ is a numerical constant of the order of $10^4$ used to scale the single protein current to the macroscopic data, $I_{n/a}$ is the current of the single protein calculated with the native/active configuration, $n_{n/a}$ is the fraction of native/active protein expected in the sample.
Of course, for a pure state, this formula reduces to the simple proportional rescaling:
\begin{equation}
I_{S}=A I_{n/a}
\label{eq:6}
\end{equation}
%
%figure7
\begin{figure}
        \centering
                \includegraphics*[width=0.45\textwidth]{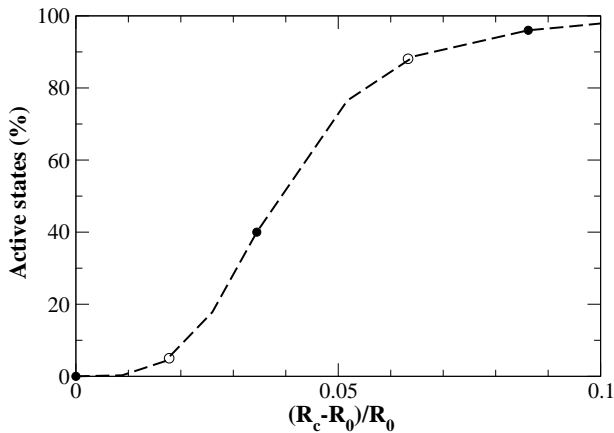}
        \caption{Per cent concentration of proteins in the active state for the different values of the interaction radius to be considered in simulations. Full circles refer to the values used  in Fig.~\ref{fig:miscele}, dashed line is the fitting obtained with Eq.~\ref{eq:5}, empty circles refer to values intermediate between the experiments reported in Ref, \cite{Jin06}, see text. }
        \label{fig:rcfit}
\end{figure}
\par
Figure  \ref{fig:rcfit}  reports the concentration of active 2NTW states in the samples vs the corresponding $R_c$ values to be used in simulations.
Symbols refer to values used in simulations and the dashed curve is a fitting obtained
from a sigmoidal Hill-like function that is commonly used in biochemistry to  describe the percentage of proteins activated by a ligand.
Its validity in fitting several  different physicochemical reactions is well known \cite{Goutelle08}, and writes:
\begin{equation}
f(x)=\frac{x^{\alpha}}{b+x^{\alpha}} \times 100,  \qquad x= \frac{R_{c}-R_{0}}{R_{0}} 
\label{eq:7}
\end{equation}
where $f$ is the percentage of proteins in the active state, 
and $R_0$=5.8 \ \AA.  
For $x^{a} =b$, half of the proteins in the sample have changed their configuration.
Here, the best fitting parameters are $a=3.91$ and $b=2.87\times 10^{-6}$, i.e. $b^{1/\alpha}=\left({6.02-R_{0}}\right)/{R_{0}}$.
The full circles reproduce the experiments reported in Fig. ~\ref{fig:rcfit}. 
For the case of the experiments in Ref. \cite{Jin06}, further binary mixtures  with the percentages suggested by Eq.~\ref{eq:7} (open circles) have been tested to to be consistent with experiments but to a less quantitative resolution of the photocurrent.

%%NEW
In the present context, the meaning of the function $f$ is the following: for an increasing number of photons impinging onto the sample, the free energy of the protein grows and, as a consequence, the value of $R_c$ also grows. 
With $R_c$  also the percentage of proteins moved to the active state grows because some of the photons hit the retinal. 
Finally, for $R_c$  larger than about $7$ \AA \, all the proteins in  the sample are in the active state. 
Further amount of photons may only improve the free energy of the protein in the active state and the internal degree of connections.
\section{Conclusions}
%
%sec_4
The paper investigates the mechanisms responsible for the photocurrent exhibited by monolayer samples of bacteriorhodopsin in the presence of an impinging green light.
To this purpose,  use is made of the INPA model implemented to account for the change of connectivity of the single protein associated with the presence of the light.
Previous results provided a satisfactory interpretation of a set of accurate measurements, performed with the c-AFM technique, in nanolayer samples and in the absence of direct light.
Accordingly, experiments were interpreted on the basis of the tertiary structure associated with the native state of the single protein.
However, a less satisfactory agreement was obtained in the region of low voltages when the same approach was applied to the case of monolayer samples of bR in the presence of light, and thus taking the  tertiary structure of the protein in its active state. 
To overcome this drawback, here we consider also  the change in the connectivity of the protein state consequent to the  enhancement of the free-energy level of the single protein induced by the presence of the light.
The increase in connectivity is accounted for by an increase of the value of the interaction radius, $R_c$, already introduced to  correlate the electrical properties with the tertiary structure of the protein.
Accordingly, the new model interprets the photocurrent using a binary mixture of results pertaining to the native and active structures with the proper values  of $R_c$ \cite{Nadia}.
Specifically, a satisfactory fit of experiments on nanolayers \cite{Jin06} is obtained by using:\, (i) for the sample in light, $R_c$ = 6.3 \ \AA \ and a binary mixture of 96\% of 2NTW and of 4 \% of 2NTU, (ii) for the sample in dark, $R_c$ = 5.8 \ \AA\ and 100  \% of 2NTU.
The c-AFM experiments, performed in the absence of direct light, are quite finely reproduced by  using a binary mixture containing up to 40\% of 2NTW and  
$R_c$=6.0 \ \AA  (see Fig.~\ref{fig:miscele}).
Therefore, the implemented model enables us to achieve a better agreement between theory and experiments in the region of low applied bias and does not modify previous findings at high values of applied bias. 
\par
We notice that the process of protein activation, in particular for opsins, is still a very open topic \cite{Onuchic97, Kobilka07} and the present approach aims to provide a further step for a better understanding of the subject.
Environmental effects, different from the presence of light, like temperature, the value of the pH, etc, should be responsible for other activation mechanisms.
Accordingly, more experiments, and structural information are necessary, and present results should give a further motivation to stimulate new experiments and formulate new theories.
Finally, this research exploits the trend in which different emerging disciplines can converge in a new branch of science, we recently introduced as proteotronics \cite{rome14}. 
Indeed, proteotronics aims to develop new devices based on the sensing properties of proteins.
In doing so, protein responses to external stimuli have chances to be better understood and used to devise biodevices of relevant importance in applied sciences.
\begin{acknowledgments}
{This research is supported by the European Commission under the Bioelectronic Olfactory Neuron Device (BOND) project within the grant agreement number 228685-2.}
\end{acknowledgments} 
%%%%%%%%%% %%%%%%%%%%%%%%%%%%%%% References

%

\begin{thebibliography}{99}
\bibitem{Lozier75}
R. H. Lozier, R. A. Bogomolni, and W. Stoeckenius, 
%(1975). 
%Bacteriorhodopsin: a light-driven proton pump in Halobacterium Halobium.
\Journal{Biophysical journal}{15}{955}{1975}.
\bibitem{Corcelli00}
A.Corcelli, M. Colella, G. Mascolo, F. P. Fanizzi, and M. Kates,
% (2000). A novel glycolipid and phospholipid in the purple membrane. 
\Journal{Biochemistry}{39}{3318}{2000}.
%
\bibitem{varo99}
G.  Varo,  L.S. Brown, R. Needleman, and J.K. Lanyi, 
%Binding of Calcium Ions to Bacteriorhodopsin, 
\Journal{Biophys. J.}{76(6)}{3219}{1999}.
%
\bibitem{etzkorn13}
M. Etzkorn \textit{et al}., 
%T.  Raschle, F. Hagn, V.  Gelev,  A.J. Rice, T. Walz, and G. Wagner1, 
%Cell-free Expressed Bacteriorhodopsin in Different Soluble Membrane Mimetics: Biophysical Properties and NMR Accessibility, 
\Journal{Structure}{21}{394}{2013}.
%
\bibitem{yoshino13}
M. Yoshino \textit{et al}.,
% T.  Kikukawa, H,  Takahashi , T. Takagi, Y. Yokoyama, H.  Amii, T. Baba, T. Kanamori, and M. Sonoyama, 
%Physicochemical Studies of Bacteriorhodopsin Reconstituted in Partially Fluorinated Phosphatidylcholine Bilayers, 
\Journal{J. Phys. Chem. B}{117}{5422}{2013}.
%
\bibitem{Gonzales00}
R.Gonz{\`a}lez-Luque, \textit{et al }.,
% M. Garavelli, F. Bernardi, M. Merch{\`a}n, M.~A. Robb, and  M. Olivucci,
% (2000). Computational evidence in favor of a two-state, two-mode model of the retinal chromophore photoisomerization.
\Journal{\PNAS}{97(17)}{9379}{2000}.
%\bibitem{lomas12}
%M.B. Lomas. PhD dissertation,
%Synthesis and applications of new biomimetic molecular switches
%dialnet.unirioja.es© El autor © Universidad de L 2012 a Rioja,
% Servicio de Publicaciones, (2013). 
%
\bibitem{subramaniam00}
S. Subramaniam and R. Henderson, 
%Molecular mechanism of vectorial proton translocation by bacteriorhodopsin, 
\Journal{Nature}{406}{653}{2000}.
%
\bibitem{Luecke00}
H.  Luecke, 
%Atomic resolution structures of bacteriorhodopsin photocycle intermediates: the role of discrete water molecules in the function of this light-driven ion pump,, 
\Journal{Biochimica et Biophysica Acta}{1460}{133}{2000}.
%
%Mallamace, F., Baglioni, P., Corsaro, C., Chen, S. H., Mallamace, D., Vasi, C., & Stanley, H. E. (2014). 
%The influence of water on protein properties. 
%J. Chem. Phys., 141(16), 165104

\bibitem{Kouyama89}
T. Kouyama and A. Nasuda-Kouyama,  
%Turnover rate of the proton pumping cycle of bacteriorhodopsin: pH and light-intensity dependences. 
\Journal{Biochemistry}{28(14)}{5963}{1989}.
\bibitem{Jin06}
Y. Jin, N. Friedman, M. Sheves M, T. He, and D. Cahen, 
%2006 Bacteriorhodopsin(bR) as an electronic conduction medium: Current transport
%through bR-containing monolayers 
\Journal{Proc. Natl. Am. Soc.}{103}{8601}{2006}
%
\bibitem{Ron10}
I. Ron \textit{et al }.,
%L. Sepunaru, S. Itzhakov, T. Belenkova, N. Friedman, I. Pecht, et al, 
%Proteins as Electronic Materials: Electron Transport through Solid-State
{J. Am. Chem. Soc.}{132}{4131}{2010}.
%
\bibitem{Sepunaru12}
L. Sepunaru, N. Friedman, I. Pecht, M. Sheves, and D.Cahen,  
%(2012). %Temperature-dependent solid-state electron transport through bacteriorhodopsin: Experimental evidence for multiple transport paths through proteins. 
\Journal{J. Am. Chem. Soc.}{134}{4169}{2012}.

\bibitem{Casuso07}
I. Casuso \textit{et al} ., 
%L. Fumagalli, J. Samitier, E. Padr{\`o}s, L. Reggiani L, V. Akimov, and G. Gomila,
% 2007 Nanoscale electrical conductivity of the purple membrane monolayer
\Journal{\PRE}{76}{041919}{2007}.
%
\bibitem{Mukho14}
S. Mukhopadhyay \textit{et al} .,
%S. R. Cohen, D. Marchak, N. Friedman, I. Pecht, M. Sheves,  and D. Cahen,  %(2014). 
%Nano-Scale Electron Transport and Photo-Dynamics Enhancement in Lipid-Depleted Bacteriorhodopsin Monomers. ACS nano.
\Journal{ACS Nano}{in press}{}{2014}.
%
\bibitem{Renu14}
V. Renugopalakrishnan \textit{et al}.,
% B. Barbiellini, C. King, M. Molinari, K. Mochalov, A. Sukhanova,  ... and S. Ramakrishna,
% (2014). Engineering a Robust Photovoltaic Device with Quantum Dots and Bacteriorhodopsin. 
\Journal{J. Phys. Chem. C.}{118}{16710}{2014}.  
%
\bibitem{patil12}
A. V. Patil, T. Premaruban, O. Berthoumieu, A. Watts, and J. J. Davis, 
\Journal{J. Phys. Chem. B}{116}{683}{2012}.
%
\bibitem{Dai14}
G. Dai, L.M. Chao,  and T. Iwasa,  
%\textit{Photocatalytic Degradation of Phenol with Bacteriorhodopsin Sensitized TiO2 Nanoparticles. In 
\Journal{Adv. Mat. Res.}{955}{415}{2014}.
%
\bibitem{Hampp00}
N. Hampp, 
%(2000). Bacteriorhodopsin as a photochromic retinal protein for optical memories. 
\Journal{Chemical Reviews}{100}{1755}{2000}.
%
\bibitem{Lima05}
S.~Q. Lima  and G. Miesen{\"o}ck, 
%. "Remote Control of Behavior through Genetically Targeted Photostimulation of Neurons". 
\Journal{Cell}{121}{141}{2005}. %doi:10.1016/j.cell.2005.02.004. PMID 15820685. edit
%\bibitem{Pastrana10}
%Pastrana, E., 
%(2010). "Optogenetics: Controlling cell function with light". 
%\Journal{Nature Methods}{8}{24}{2010}.
% doi:10.1038/nmeth.f.323.
%
\bibitem{Deisseroth10}
K. Deisseroth,
%(2010). "Optogenetics". 
\Journal{Nature Methods}{8}{26}{2010}. 
%doi:10.1038/nmeth.f.324. PMID 21191368.
%Proteotronics
%
%\bibitem{karplus11}
%M. Karplus, \Journal{Nature Chemical Biology}{7}{401}{2011}.
%
%
%\bibitem{Luecke00}
%H. Luecke, Biochim. Biophys. Acta 1460 133-156 (2000) 
%
\bibitem{Wickstrand14}
 C. Wickstrand,  R. Dods,  A. Royant,  and R. Neutze, 
% Bacteriorhodopsin: Would the real structural intermediates please stand up?. 
{Biochimica et Biophysica Acta(BBA)-General Subjects}{}{}{(2014)}.

%
\bibitem{Berman00}
 H.~M. Berman \textit{et al}.,
 %J. Westbrook, Z. Feng, G. Gilliland, T.~N. Bhat, H. Weissing, 
 %I.~N. Shindyalov, and P. Bourne,
 % 2000 The protein data bank 
 \Journal{Nucleic Acids Research}{28}{235}{2000}.
%
\bibitem{Lanyi06}
J. K. Lanyi and B. Schobert,
\Journal{J Mol Biol.}{365}{1379-92}{2007}.
%. Epub 2006 Nov 10.Structural changes in the L photointermediate of bacteriorhodopsin.
%
\bibitem{Alfinito11c}
E. Alfinito, J.~-F. Millithaler, and L. Reggiani,
% 2011 Charge transport in purple membrane monolayers:
%A sequential tunneling approach \
\Journal{\PRE}{83}{042902}{2011}.
%
\bibitem{Alfinito09}
E. Alfinito and L. Reggiani,
%2009 Charge transport in bacteriorhodopsin monolayer 
\Journal{Europhys. Lett.}{85}{68002}{2009}.
%
\bibitem{Alfinito09e}
E. Alfinito, C. Pennetta,  and L. Reggiani, 
%{\it Smell nanobiosensors: Hybrid systems based on the electrical response to odorant capture theory and experiment},
\Journal{AIP Conference Proceedings}{1137}{115}{2009}. 
%
\bibitem{Alfinito09d}
E.Alfinito and L. Reggiani, 
%(2009b). Detecting conformational change by current
%transport in proteins: The case of bacteriorhodopsin monolayers, 
\Journal{J. Phys: Conf. Ser.}{193(1)}{012107}{2009}.
%of Physics: Conference Series 193, 1, p. 012107, URL http://stacks.iop.
%org/1742-6596/193/i=1/a=012107.

%
%

%
%
\bibitem{Nadia} 
E. Alfinito, J.~F.Millithaler, L. Reggiani, N. Zine, and N. Jaffrezic-Renault,  %(2011). Human olfactory receptor 17-40 as an active part of a nanobiosensor: a microscopic investigation of its electrical properties. 
\Journal{RSC Advances}{1}{123}{2011}.
\bibitem{Alfinito13b}
E. Alfinito, J. Pousset, L. Reggiani, and K. Lee, 
%Photoreceptors for a light biotransducers: ...
\Journal{Nanotechnology}{24}{39551}{2013}.
%
\bibitem{Alfinito13c}
E. Alfinito and L. Reggiani, 
%Evidence of Gumbel distributions of conductance fluctuations in bacteriorhodopsin thin-films , 
\Journal{\JPC}{25}{375103}{2013}.
%
\bibitem{Alfinito14a}
E.Alfinito and L. Reggiani,
% Opsin vs opsin: New materials for biotechnological applications,
\Journal{\JAP}{ 116 }{064901}{2014}.
%
\bibitem{Alfinito14b}
E. Alfinito, J. Pousset, and L. Reggiani, 
\Journal{J. Phys. Conf. Ser.}{490} {012134}{2014}.
%%%%%%%%%%%%%%%%%%%%%%%%%%%%%%%%
 
%
 \bibitem{rome14}
 E. Alfinito,  L. Reggiani, and J. Pousset, 
 %\textit{Proteotronics: Electronic devices based on proteins }, 
 cond-mat 1405.3840;
 E. Alfinito, J. Pousset, and L. Reggiani
 \textit{Protein-Based Electronics: Transport Properties and Application. Towards the Development of a Proteotronics} (Pan Stanford Publishing Pte. Ltd.
Penthouse Level, Suntec Tower 3
8 Temasek Boulevard
Singapore 038988, in press)
%
\bibitem{Alfinito05}
E. Alfinito, V. Akimov, C. Pennetta, L. Reggiani, and G. Gomila, 
%{\it Thermal fluctuations of a gpcr: A two force constant model},
\Journal{AIP Conference Proceedings}{800}{381}{2005}.
%, doi:http://dx.doi.org/
%10.1063/1.2138641, URL http://scitation.aip.org/content/aip/
%proceeding/aipcp/10.1063/1.2138641
%
\bibitem{Akimov06}
V. Akimov \textit{et al} .,
%E. Alfinito, C. Pennetta, L. Reggiani, J. Minic, T. Gorojankina, E. Pajot-
%Augy, and R. Salesse, 
%{\it An impedance networkmodel for the electrical  properties of a single-protein nanodevice}, 
in \textit{Nonequilibrium Carrier Dynamics in Semiconductors}, edited by M. Saraniti and U. Ravaioli ( Springer Proceed-
ings in Physics, 2006) Vol. 110 
%(Springer Berlin Heidelberg), 
7, pp. 229-232.

%
\bibitem{Alfinito08}
E. Alfinito, C. Pennetta, and L. Reggiani,
% 2008 A network model to correlate conformational change and the impedance spectrum of single proteins  
 \Journal{Nanotechnology}{19}{065202}{2008}
%
%
%
\bibitem{Alfinito09a}
E. Alfinito, C. Pennetta,  and L. Reggiani, L. 
%Topological change and impedance spectrum of rat olfactory receptor i7: A comparative analysis with bovine rhodopsin and bacteriorhodopsin, 
\Journal{\JAP}{105(8)}{084703}{2009}.
 % 
\bibitem{Simmons63}
   J.~G. Simmons,
  %1963 Generalized formula for the electric tunnel effect between similar electrodes separated by a thin insulating film 
  \Journal{\JAP}{34}{1793-1803}{1963}.
%  
\bibitem{Onuchic97}
J. N. Onuchic, Z. Luthey-Schulten, and P. G.  Wolynes, 
%(1997). Theory of protein folding: the energy landscape perspective. 
\Journal{Ann. Rev. Phys. Chem.}{48}{545}{1997}.
%\bibitem{Onuchic04}
%Onuchic, J. N., & Wolynes, P. G. (2004). Theory of protein folding. Current opinion in structural biology, 14(1), 70-75.
%
 \bibitem{Kobilka07}
B. K.  Kobilka, and X. Deupi,
% (2007). Conformational complexity of G-protein-coupled receptors. 
\Journal{Trends in pharmacological sciences}{28}{397}{2007}.
%
\bibitem{Goutelle08}
S. Goutelle, M. Maurin, F. Rougier, X. Barbaut, L. Bourguignon, M. Ducher, and P. Maire,  %(2008). The Hill equation: a review of its capabilities in pharmacological modelling. 
%, 22(6), 633-648.
% (2007). Conformational complexity of G-protein-coupled receptors. 
\Journal{Fundamental \& clinical pharmacology}{22}{633}{2008}.
%\bibitem{Zvyargin07}
%I. Zvyargin, 
% Charge transport via delocalized states in disordered materials
%in: \textit{Charge transport in disordered solids} ed S Baranovski (J. Wiley \& Sons, The Atrium, Southern gare, Chichester, England, 2006) 1-48 
\end{thebibliography}
\end{document}